\documentclass[aps,prl,twocolumn,groupedaddress,showpacs]{revtex4}
\usepackage{psfig,graphics,graphicx}
\usepackage{color}
\usepackage{amssymb}
\usepackage{bookmath}
\usepackage{bm}
\begin{document}
\title{Nodal structure of quasi-2D superconductors probed by magnetic field}
\author{A. Vorontsov}
\author{I. Vekhter}
\affiliation{Department of Physics and Astronomy,
             Louisiana State University, Baton Rouge, Louisiana, 70803, USA}
\date{\today}

\begin{abstract}
We consider a quasi two-dimensional superconductor with line nodes
in an in-plane magnetic field, and compute the dependence of the
specific heat, $C$, and the in-plane heat conductivity, $\kappa$,
on the angle between the field and the nodal direction in the
vortex state. We use a variation of the microscopic
Brandt-Pesch-Tewordt method that accounts for the scattering of
quasiparticles off vortices, and analyze the signature of the
nodes in $C$ and $\kappa$. At low to moderate fields the specific
heat anisotropy changes sign with increasing temperature.
Comparison with measurements of $C$ and $\kappa$ in CeCoIn$_5$
resolves the contradiction between the two in favor of the
$d_{x^2-y^2}$ gap.
\end{abstract}
\maketitle

\paragraph{Introduction.} Remarkable properties of
superconductors below the transition temperature, $T_c$, are due
to formation of the Cooper pairs and opening of the energy gap in
the single particle spectrum. The structure of the gap is
intimately related to the nature of the pairing interaction, and
therefore its experimental determination is of much importance. In
simple metals the phonon-mediated electron attraction and the gap
are isotropic around the Fermi surface. In contrast, in the
majority of recently discovered superconductors  the gap is
strongly anisotropic, and often vanishes (has nodes) for selected
directions in the momentum space.

While the existence and topological structure (line vs. point) of
the nodes can be inferred from the power laws in the temperature
dependence of the thermodynamic and transport properties,
experimental determination of the nodal {\em directions} is more
challenging. Currently the most widely used probes utilize an
applied magnetic field, $\bm H$, as a directional probe of
unpaired quasiparticles (QP).

Low energy quasiparticles only exist in the near-nodal regions.
They are excited by temperature, $T$, at all nodes, and by the
field, predominantly at the nodes where the QP velocity is normal
to $\bm H$. Consequently the QP density changes with the relative
orientation of the field with respect to the nodes. It was
predicted that the low-energy density of states (DOS), and the
electronic specific heat, $C$, have minima when the magnetic field
is aligned with the nodes \cite{IVekhter:1999R}.

Due to challenges in measuring the electronic contribution to
$C(T,H)$, so far experimental results exist for few materials
\cite{TPark:2003,HAoki:2004}. At the same time the anisotropy of
the electronic thermal conductivity, $\kappa_{xx}\equiv\kappa$,
under a rotating field was determined in several systems
\cite{FYu:1995,KIzawa:other,KIzawa:CeCoIn5}. The orientational
dependence of the in-plane $\kappa(\bm H)$ is more complex than
that of the specific heat as it combines the anisotropy due to
nodal structure with that due to the difference in scattering
normal to and parallel to the vortices. Theoretical interpretation
of the anisotropy of transport properties proved elusive as the
existing theories have difficulties accounting for the two
modulations, and cannot unequivocally determine whether a local
minimum or a maximum occurs when the field is along a node.
Consequently, more theoretical work is needed to examine the
experimentally conjectured gap anisotropy
\cite{KIzawa:other,KIzawa:CeCoIn5}. Moreover, in heavy fermion
CeCoIn$_5$, where data on both $C$ and $\kappa$ anisotropy exist,
the conclusions appear contradictory: $d_{xy}$ ($d_{x^2-y^2}$) gap
symmetry was inferred from $C$ ($\kappa$)
\cite{HAoki:2004,KIzawa:CeCoIn5}.

In most approaches the effect of ${\bm H}$ on the QPs is included
via the Doppler energy shift due to supercurrents around the
superconducting vortices \cite{GVolovik:1993}, and the QP lifetime
is not affected directly \cite{CKubert:1998}. In reality, vortices
scatter quasiparticles carrying the heat current and not just
shift their energy.  Consequently, the behavior of $\kappa(T,\bm
H)$ is determined by the competition between the enhancement of
the DOS and the vortex scattering. There are indications that
vortex scattering affects transport already at moderate, compared
to the upper critical field $H_{c2}$, fields
\cite{IVekhter:1999,ACDurst:2003}, but its effect for the in-plane
field and on the anisotropy in $C(\bm H)$ and $\kappa(\bm H)$ has
not been studied.

In this Letter we present a unified microscopic approach to
computing the anisotropy of thermodynamic and transport properties
of nodal superconductors in the vortex state at moderate to high
magnetic field. We apply the method to quasi two-dimensional
superconductors with line nodes, and consider the field rotated in
the $xy$ plane. We are able to account for both twofold and
fourfold variations of $\kappa(T,\bm H)$ with angle. For
$d_{x^2-y^2}$ gap, the competition between the {\em transport}
scattering rate and the DOS leads to switching from minima to
maxima for $\bm H$ along the nodes in the fourfold part of
$\kappa(T,\bm H)$ upon increasing temperature. We find that, due
to the field dependence of the {\em single particle} lifetime, the
minima and the maxima in the specific heat also switch at higher
$T$ and $H$. Hence in a wide $T$-$H$ range the maxima (rather than
minima) in $C(T,\bm H)$ indicate a nodal direction. Our results
for the $d_{x^2-y^2}$ gap are in a semi-quantitative agreement
with experiment on CeCoIn$_5$.

\paragraph{Model and approach.} We consider a quasi-two
dimensional system with a model Fermi surface (FS) $p_f^2 = p_x^2
+ p_y^2 - (r^2 \, p_f^2) \cos (2 s\, p_z/r^2 p_f)$. The
parameters, $r$ and $s$, determine the corrugation amplitude along
the $z$-axis and the ratio $v_{f,z}/v_{f,\perp}$ (at
$p_x^2+p_y^2=p_f^2$) respectively. We assume a separable pairing
interaction,
$V(\hat{\vp},\hat{\vp}^\prime)=V_0\cY(\hat{\vp})\cY(\hat{\vp}^\prime)$,
where $\cY(\hat{\vp})$ are the normalized basis functions for the
angular momentum eigenstates, so that $\cY(\hat{\vp})=\sqrt{2}
\cos 2\phi_{\hat{p}}$ ($\cY(\hat{\vp})=\sqrt{2} \sin
2\phi_{\hat{p}}$) for $d_{x^2-y^2}$ ($d_{xy}$) gap. Here
$\phi_{\hat{p}}$ is the angle between the projection of the
momentum $\vp$ onto the $xy$ plane and the $x$-axis. The magnetic
field, $\bm H$, is in the $xy$ plane, at an angle $\phi_0$ to the
$x$-axis. We ignore Zeeman splitting, which is justified for most
systems, see below.

We rotate $x$ and $y$ axes around $z$ to choose the field
direction as the new $x$-axis, and model the spatial dependence of
superconducting order parameter by an Abrikosov-like solution,
$\Delta(\vR, \hat{\vp}) = \Delta(\vR) \, \cY(\hat{\vp})$, where
$\Delta(\vR) = \sum_n \, \Delta_n \, \braket{\vR}{n}$. Here
$\braket{\vR}{n} = \sum_{k_y} \, C^{(n)}_{k_y} \, \Phi_n(
(z-\Lambda^2 k_y)/\sqrt{s} \Lambda) \, \exp(ik_y y)$, $\Phi_n(z)$
is the eigenfunction of the $n$-th state of a harmonic oscillator,
$\Lambda^2 = \hbar c/2|e|H$, and normalized $C^{(n)}_{k_y}$
($\sum_{k_y} |C^{(n)}_{k_y}|^2=1$) determine the structure of the
vortex lattice.  The factor $\sqrt{s}$ ensures that the functions
$\Phi_n$ are the solutions to the linearized equations for the
order parameter \cite{HKusunose:2004,AVorontsov:2006}, and
$\Delta_n$ is the amplitude of the $n$th spatial component. In
s-wave superconductors the vortex state is a superposition of
$n=0$ states only; however, for singlet pairing in the higher
angular momentum channels the self-consistency requires admixture
of other even $n$ \cite{ILukyanchuk:1987,HAdachi:2005}. In our
calculations we use $n=0,2,4$, since inclusion of higher $n$ does
not alter the results.

We use the quasiclassical Green's function approach
\cite{GEilenberger:1968}, where the equation for the retarded
(index $R$) anomalous (Gor'kov) Green's function, is given by
    \begin{eqnarray} \left[-2i\tilde{\vare} + \vv_f(\hat{\vp}) \left(
\gradR - i \frac{2|e|}{\hbar c} \vA(\vR) \right) \right] \,
f^R(\hat{\vp}, \vR; \vare)
\nonumber \\
 = 2 \tilde{\Delta} \, i g^R(\hat{\vp}, \vR; \vare) \,.
 \label{Feq}
    \end{eqnarray}
Here $\tilde{\vare} = \vare - \Sigma^R(\vR; \vare)$,
$\tilde{\Delta} = \Delta(\hat{\vp}, \vR) + \Delta^R_{imp}$, and we
treat the self-energies $\Sigma^R$, $\Delta^R_{imp}$ due to
impurity scattering in the self-consistent $T$-matrix
approximation, and focus here on the unitarity limit (scattering
phase shift $\pi/2$) for clean systems (normal state scattering
rate $\Gamma\ll T_c$). The normal electron Green's function
$g^R(\hat{\vp}, \vR; \vare)$ is determined via the normalization
condition $(g^R)^2 - \ul{f}^R f^R = - \pi^2$, where
$\ul{f}^R(\hat{\vp}, \vR; \vare) = f^R(-\hat{\vp}, \vR;
-\vare)^*$. Eq.(\ref{Feq}) is complemented by the self-consistency
condition for $\Delta$.

To solve the quasiclassical equations we make use of the
approximation due to Brandt, Pesch, and Tewordt (BPT)
\cite{BPT:1967}, and replace  $g^{R}(\bm R,\hat{\bm
p},\varepsilon)$, by its spatial average. If $\bm K$ is a vector
of the reciprocal vortex lattice, the Fourier components
$g^{R}(\bm K) \propto\exp(-\Lambda^2 {\bm K}^2)$, hence the
spatial average $\bm K=0$ dominates\cite{BPT:1967}. The method is
nearly exact at $H\lesssim H_{c2}$, but gives semi-quantitatively
correct results down to much lower fields. In extreme type-II
\cite{EHBrandt:1976}, and in the nodal superconductors
\cite{HWon:1996,IVekhter:1999} the method remains valid over
almost the entire range $H_{c1}\ll H<H_{c2}$ \cite{comment1}, and
was used to study unconventional superconductors in the vortex
state \cite{IVekhter:1999,HKusunose:2004,LTewordt:2005}.

With averaged $g^{R}$ we solve Eq.(\ref{Feq}) by expanding
$\Delta, f^R, \ul{f}^R$ in the orthonormal set
$\{\braket{\vR}{n}\}$, and using the ladder operators for the
oscillator states, $a$ and $a^\dagger$,
\cite{AHoughton:1998,HKusunose:2004} to rewrite $\vv_f(\hat{\vp})
\left( \gradR - i (2|e|/ \hbar c) \vA(\vR) \right)=v_-
a-v_+a^\dagger$, where $v_\pm=[v_z/\sqrt s\pm i v_y\sqrt
s]/(\Lambda\sqrt 2)$. Lengthy but straightforward calculation
gives the closed form expressions for functions $f^R, \ul{f}^R$,
and $g^{R}$ for a set $\Delta_n$, which is then determined
self-consistently. For $n=0$ only we find

\begin{equation}
g^R(\hat\vp; \vare) = \frac{-i \pi}{
\sqrt{1-i\sqrt{\pi}\left(\frac{2\Lambda\Delta_0}{|\tilde{v}_f^\perp|}\right)^2
\cY^2(\hat\vp) \, W^\prime(
\frac{2\tilde{\vare}\Lambda}{|\tilde{v}_f^\perp|} )} } \,
\label{eq:dos}
\end{equation}
similar to Refs.\cite{IVekhter:1999,HKusunose:2004}. Here $W(z) =
\exp(-z^2)\,\mbox{erfc}(-iz)$, and the dependence on the field
direction is via the rescaled component of the Fermi velocity
normal to $\bm H$, $|\tilde{v}_f^\perp(\vp_f)|^2=v_{f,z}^2/s+ s
v_{f,y}^2$. The expression for arbitrary $n$ has a similar
structure, but involves combinations of $\Delta_n$ and higher
derivatives of $W(z)$  \cite{AVorontsov:2006}.

The density of states (DOS) is $N(T, H; \vare)/N_f =
(-1/\pi)\,\int d\hat{\vp}_f \,  \Im\, g^R(\hat{\vp}; \vare)$,
where $N_f$ is the normal state DOS per spin at the Fermi level.
From the DOS we determined the entropy, $S(T,\bm H)$, computed the
specific heat, $C = T \,\partial{S}/\partial{T}$, by numerical
differentiation, and verified that far from the transition
$T_c(H)$ the expression
\begin{equation}
  C(T,\bm H) = \frac{1}{2}  \int\limits^{+\infty}_{-\infty}
\; d\vare \; \frac{\vare^2 \; N(T, \bm H; \vare)}{T^2
\cosh^2(\vare/2T)} \,, \label{Eq:C}
\end{equation}
gives accurate results. We use Eq.(\ref{Eq:C}) below.

The thermal conductivity tensor can be obtained using Keldysh
technique \cite{AVorontsov:2006}, and has a particularly simple
form for the Born and unitarity scattering limits
\cite{PJHirschfeld:1988,MJGraf:1996}
\begin{eqnarray}
&&\frac{\kappa_{ij}(T,H)}{T} = \int\limits^{+\infty}_{-\infty} \;
\frac{d\vare}{2 T} \frac{\vare^2}{T^2} \cosh^{-2}\frac{\vare}{2T}
\\
\nonumber &&\times \int \, d\hat{\vp} \, v_{f,i} v_{f,j} \, N(T,
\bm H; \hat{\vp}, \vare) \; \tau_H(T, H; \hat{\vp},\vare) \,.
\end{eqnarray}
Here $N$ is the angle-dependent DOS, and $\tau_H$ has the meaning
of the {\em transport} lifetime due to both impurity and vortex
scattering.  For $n=0$ channel
\cite{PKlimesch:1978,IVekhter:1999,HKusunose:2004},
\[
\nonumber
  \frac{1}{2\tau_H} =
- \Im \Sigma^R + {2\sqrt{\pi} \Lambda \Delta_0^2 \cY^2 \over
|\tilde{v}_f^\perp|} \frac{\Im[g^R \, W(
2\tilde{\vare}\Lambda/|\tilde{v}_f^\perp|)]}{\Im \, g^R} \,,
\]
and addition of other channels results in a more complex
combination of the $W$-function and its derivatives
\cite{AVorontsov:2006}.

\begin{figure}[t]
\centerline{\includegraphics[height=6cm]{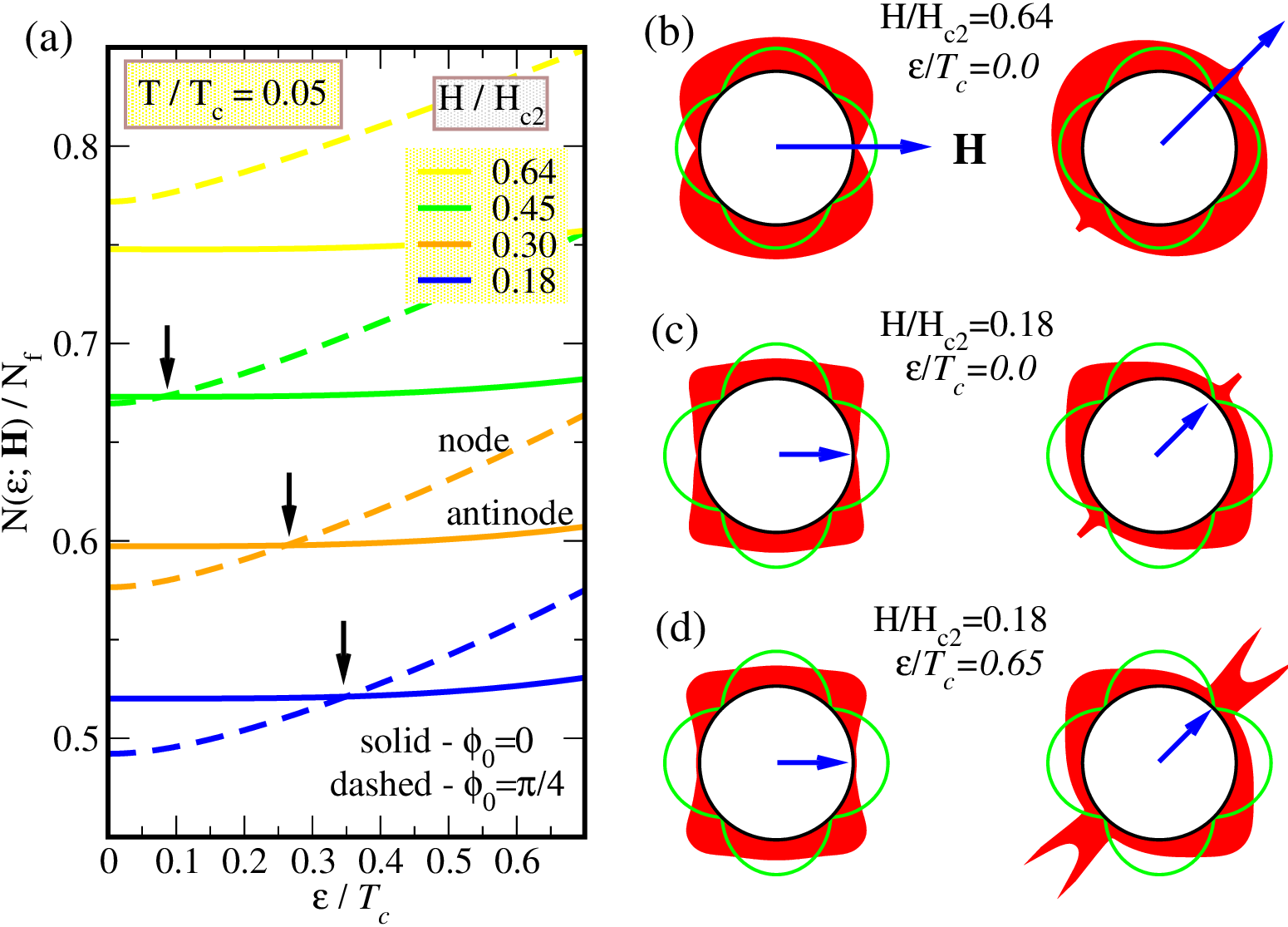}}
    \caption{(a)
Low energy DOS for $\bm H\| antinode$ and $\bm H\| node$  at
$T=0.05\, T_c$. Arrows indicate $\vare(H)$, see text. (b)--(d)
Angle-resolved DOS plotted at each position on the FS. Net DOS is
proportional to the red shaded area, see text.} \label{fig:dos}
\end{figure}

\paragraph{Results.} We show the results for $s=0.5$, $r=0.5$,
which gives $H_{c2}^{ab}/H_{c2}^{c}\approx 2.45$ similar to
CeCoIn$_5$, and for $\Gamma/(2\pi T_c)=0.007$ \cite{comment2}. The
main qualitative difference from previous results is already clear
from the behavior of the DOS shown in Fig.\ref{fig:dos}(a). In the
Doppler shift method the DOS anisotropy, $N(\vare,\bm H\|
antinode)-N(\vare,\bm H\| node)$, increases as $\sqrt H$ at
$\vare=0$, and vanishes at $\vare\sim v_F\sqrt s/\Lambda$
~\cite{IVekhter:1999R,IVekhter:2001}. In contrast, the anisotropy
in the residual ($\vare=0$) DOS has a maximum at $H\sim 0.1
H_{c2}$, and {\em reverses} above the field $H\sim 0.5 H_{c2}$.
Below this field the anisotropy also {\em changes sign} at a
finite $\vare (H)$, see Fig.\ref{fig:dos}(a).

Since $W(z)$ and its derivatives in Eq.(\ref{eq:dos}) are complex
functions, our $N(\vare)$ cannot be obtained from the BCS DOS by a
simple energy shift: there is an anisotropic single particle
scattering rate due to scattering from vortices (vanishingly small
along $\bm H$, largest normal to $\bm H$). This occurs since in
the BPT method $g^R$ is averaged incoherently in different unit
cells of the vortex lattice. Vortex scattering is pairbreaking,
and hence enhances the DOS.

Consider the contribution to the DOS from different FS regions and
focus first on $\vare=0$, Fig.\ref{fig:dos}(b),(c). At low fields
the vortex scattering rate is small, and the unpaired states
emerge only near the nodes. When $\bm H\| node$ the number of such
states at the nodes aligned with the field is small, while $\bm
H\| antinode$ produces states at all nodes, see
Fig.\ref{fig:dos}(c), and $N(0,\bm H\| antinode)>N(0,\bm H\|
node)$  as in the Doppler shift method \cite{IVekhter:1999R}. At
higher fields the pairbreaking is stronger, and, for $\bm H\|
antinode$, it generates the unpaired states in all directions on
the FS {\em except} close to the field direction, see
Fig.\ref{fig:dos}(b). For $\bm H\| node$ the field-induced states
appear everywhere on the FS [ Fig.\ref{fig:dos}(b)], leading to
the anisotropy reversal \cite{PMiranovic:2003}.

Now consider $0<\vare\ll\Delta_0$, Fig.\ref{fig:dos}(d). In a pure
system at zero field the dominant contribution to $N(\vare)$ is
from sharp peaks at directions $\hat{\bf p}_\vare$, close to the
nodes, such that $|\Delta(\hat{\vp}_\vare)| = \vare$.  Scattering
redistributes the spectral weight from the peak. Vortex scattering
at $\hat\vp_\vare$, increases with $H$, and is stronger for $\bm
H\perp \hat\vp_\vare$  than for $\bm H\| \hat\vp_\vare$. For
$\vare>\vare(H)$ $\bm H\|node$ fills the near-nodal states, but
does not broaden the peak nearest to $\hat\vp_\vare$
significantly, Fig.\ref{fig:dos}(d).  For $\bm H\| antinode$ most
of the weight in the peak is shifted away. This leads to a higher
DOS at $\vare>\vare(H)$ for the field along the node, see
Fig.\ref{fig:dos}(a).

\begin{figure}[t]
\centerline{\includegraphics[height=6cm]{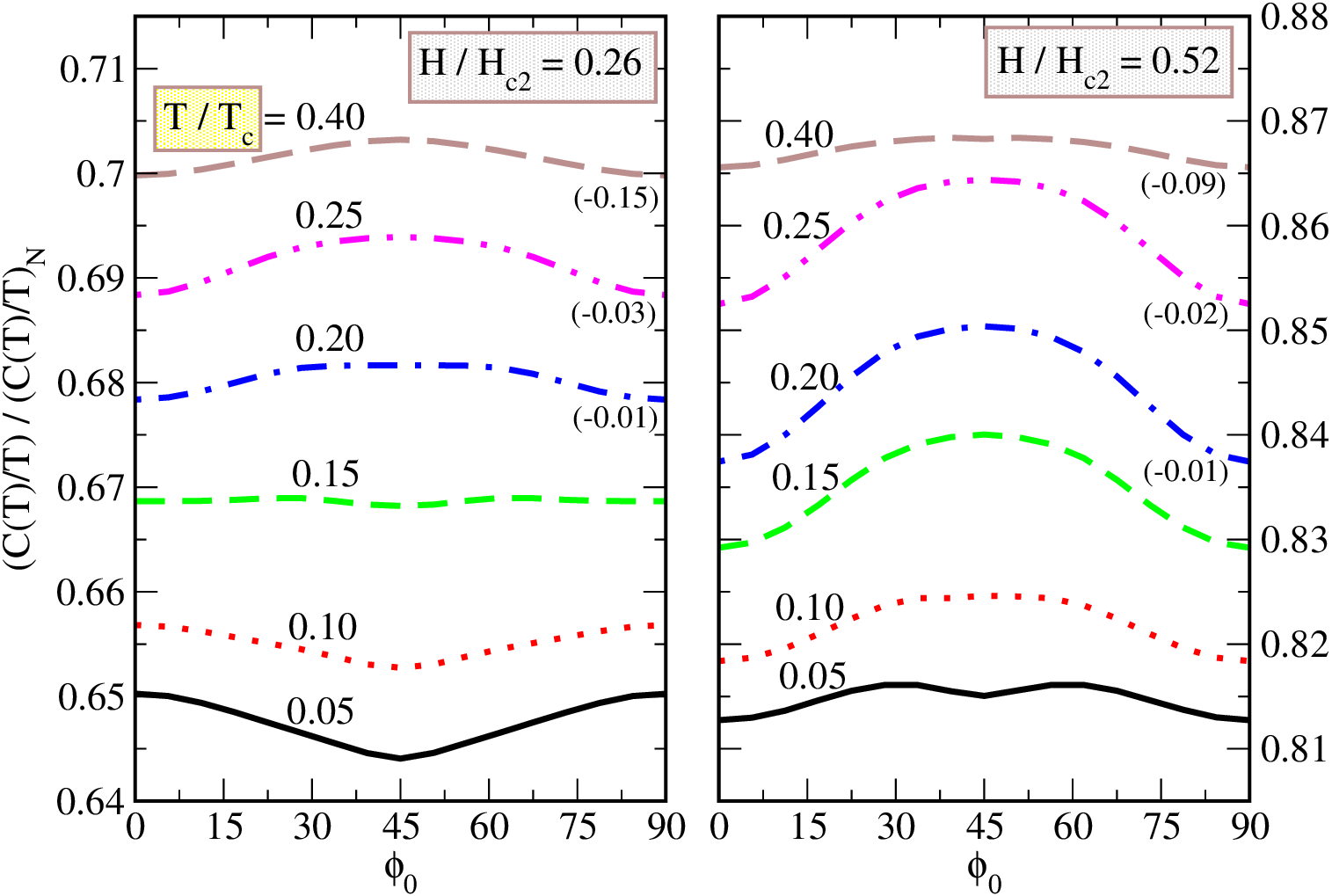}}
\caption{Specific heat anisotropy for different $T$ at two fields.
Nodal direction $\phi_0=45^\circ$. Left panel: $H=0.26H_{c2}$,
inversion of the anisotropy occurs at $T_0\sim 0.15T_c$. Right
panel: $H=0.52 H_{c2}$, no minimum for the field along the node at
any $T$. Some curves are shifted by the amount indicated in
parentheses.} \label{fig:hcap}
\end{figure}

Fig.~\ref{fig:hcap} shows the specific heat anisotropy. The
dominant contribution to $C(T,\bm H)$ is from the DOS at energies
$\varepsilon\sim 2.4 T$, see Eq.(\ref{Eq:C}). Therefore even at
low fields, when $N(0,\bm H\| antinode)>N(0,\bm H\| node)$, the
anisotropy of the specific heat is reversed at $T_0\sim
0.1$-$0.2\,T_c$ \cite{T0}. Below $T_0$, $C(T,\bm H)$ has a minimum
when the field is applied along a nodal direction, while at higher
$T$ it has a {\em maximum} for this field orientation. At higher
fields ${\bm H}$ along a node gives a maximum in the
angle-dependent specific heat at all but the lowest $T$. This
result directly affects the experimental determination of the
nodal directions.

The measurements on CeCoIn$_5$ were carried out for $0.18\leq
H/H_{c2}\leq 0.5$, and for $T>0.1 T_c$ \cite{HAoki:2004}. For this
system $H_{c2}$ is Pauli limited \cite{bia03}, and the orbital
critical field, $H_{c2}^{(orb)}$, may be as high as $2.5H_{c2}$.
We find that $T_0(H)$ in Fig. 2 is weakly field dependent for
$0.1\leq H/H_{c2}^{(orb)}\leq 0.3$. Therefore our results indicate
that maxima, rather than minima in $C(T,\bm H)$ occur for $H\|
node$ in this regime, and the data of Ref.~\onlinecite{HAoki:2004}
support the $d_{x^2-y^2}$ gap symmetry (rather than the $d_{xy}$
order inferred by the authors from the low $T$, low $H$ theory
\cite{IVekhter:1999R}). While the BPT approach likely
overestimates the vortex scattering at low $H/H_{c2}$, the
extended range of this shape of $C(T,\bm H)$ is also in favor of
$d_{x^2-y^2}$ pairing.

\begin{figure}
\centerline{\includegraphics[height=6cm]{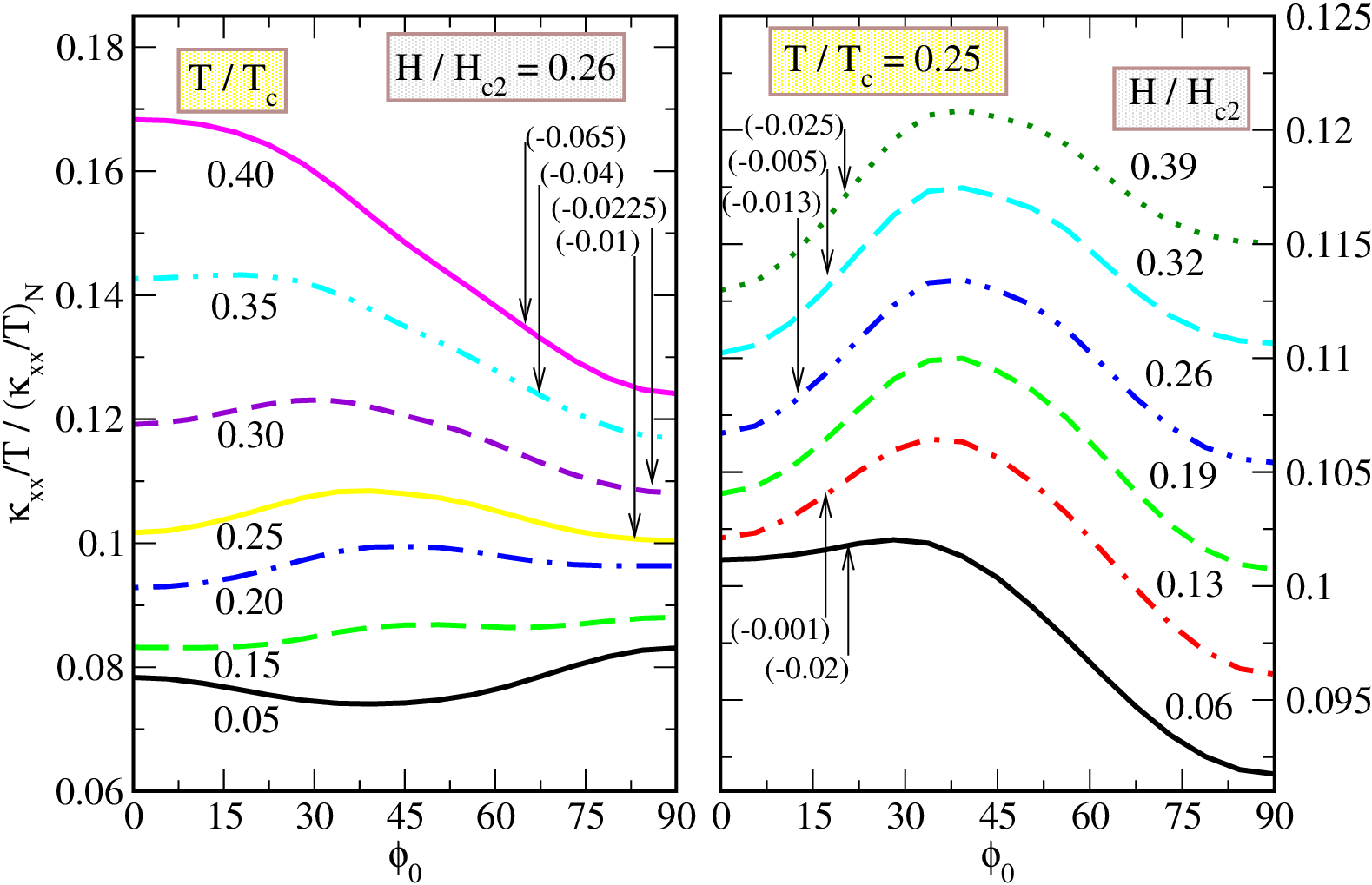}} \caption{
Thermal conductivity anisotropy  for different  $T$ at a
$H=0.26H_{c2}$ (left panel), and  for different $H$ at $T=0.25T_c$
(right panel). Curves are shifted for clarity by the amount
indicated. Left panel: notice the reversal of the
$0^\circ-90^\circ$ anisotropy and the change from minimum to
maximum at $45^\circ$. Right panel provides direct comparison with
Ref.\onlinecite{KIzawa:CeCoIn5}.} \label{fig:hconTscan}
\end{figure}

This conclusion is supported by the analysis of the thermal
conductivity, shown in Fig.~\ref{fig:hconTscan} for the heat
current $\bm j_q\| \hat{\bm x}$. Comparison can be made with
$\kappa(T,\bm H)$ as shown in Ref.~\onlinecite{KIzawa:CeCoIn5} for
$T\leq 0.5T_c$ and fields $H\leq 0.3 H_{c2}$. In this range we
find the overall shape of the curve and the amplitude of the peak
($\sim 1$-$3$\%) at the nodal angle always similar to those in the
right panel Fig.~\ref{fig:hconTscan}. We also confirmed that
$d_{xy}$ gap is inconsistent with the results of
Ref.~\onlinecite{KIzawa:CeCoIn5}. The orientational dependence of
$C(T,\bm H)$ for the $d_{xy}$ gap is obtained by a 45$^\circ$
degree rotation in Fig.\ref{fig:hcap}. In contrast, the variation
of $\kappa(T,\bm H)$ with angle is different for the two cases
since the heat current $\bm j_q\|\hat{\bm x}$ is along the
antinodal (nodal) direction for the $d_{x^2-y^2}$ ($d_{xy}$) gap
\cite{AVorontsov:2006}.

The fourfold angle dependence of $\kappa(\bm H)$ is superimposed
on the twofold variation due to relative orientation of $\bm j_q$
and $\bm H$. The field is strongly pairbreaking for $\hat{\bf
p}\perp\bm H$, and QPs which contribute the most to $\kappa(T,\bm
H)$, are created for $\bm H\perp \bm j_q$ \cite{KMaki:1967}.
Consequently, at low $T$, $\kappa(T,\bm H\perp\bm
j_q)>\kappa(T,\bm H\|\bm j_q)$. At higher $T$ the QP are thermally
excited, and $\kappa(T,\bm H\perp\bm j_q)<\kappa(T,\bm H\|\bm
j_q)$ due to vortex scattering, see Fig.~\ref{fig:hconTscan}.
$\kappa(T,{\bm H}\|node)$ has a local minimum at low $T,H$, but
develops a maximum at higher $T$. At yet higher $T$, $\kappa(T,\bm
H)$ is essentially twofold.

{\it Conclusions.} We developed a fully self-consistent
microscopic calculation of the magnetic field induced anisotropy
of thermal and transport properties of nodal superconductors. We
applied the method to quasi-two dimensional systems with line
nodes. We 1) found that, while at very low temperatures and fields
the specific heat anisotropy is in agreement with the Doppler
shift analysis, at higher $T$ and $H$ it changes sign, and, over a
wide range of fields and temperatures, exhibits a maximum, rather
than a minimum, for the field applied along a nodal direction; 2)
accounted simultaneously for the two-fold and fourfold pattern in
the angle-dependence of the thermal conductivity; 3) used the
approach to resolve a controversy regarding the symmetry of the
order parameter in CeCoIn$_5$ in favor of $d_{x^2-y^2}$ gap. We
believe that our approach provides a reliable tool to be used
alongside experimental measurements for determination of the nodal
directions in novel superconductors.

This work was partly done at KITP with support from NSF Grant
PHY99-07949;  and was also supported by the Board of Regents of
Louisiana. We thank D.~A.~Browne, C.~Capan, P.~J.~Hirschfeld,
Y.~Matsuda, and T.~Sakakibara for discussions.

\bibliographystyle{apsrev}


\begin{thebibliography}{36}
\expandafter\ifx\csname
natexlab\endcsname\relax\def\natexlab#1{#1}\fi
\expandafter\ifx\csname bibnamefont\endcsname\relax
  \def\bibnamefont#1{#1}\fi
\expandafter\ifx\csname bibfnamefont\endcsname\relax
  \def\bibfnamefont#1{#1}\fi
\expandafter\ifx\csname citenamefont\endcsname\relax
  \def\citenamefont#1{#1}\fi
\expandafter\ifx\csname url\endcsname\relax
  \def\url#1{\texttt{#1}}\fi
\expandafter\ifx\csname
urlprefix\endcsname\relax\def\urlprefix{URL }\fi
\providecommand{\bibinfo}[2]{#2}
\providecommand{\eprint}[2][]{\url{#2}}

\bibitem[{\citenamefont{Vekhter et~al.}(1999)\citenamefont{Vekhter, Hirschfeld,
  Nicol, and Carbotte}}]{IVekhter:1999R}
\bibinfo{author}{\bibfnamefont{I.}~\bibnamefont{Vekhter}} et al.,
  \bibinfo{journal}{Phys. Rev. B} \textbf{\bibinfo{volume}{59}},
  \bibinfo{pages}{R9023 } (\bibinfo{year}{1999}).

\bibitem[{\citenamefont{Park et~al.}(2003)\citenamefont{Park, Salamon, Choi,
  Kim, and Lee}}]{TPark:2003}
\bibinfo{author}{\bibfnamefont{T.}~\bibnamefont{Park}} et al.,
  \bibinfo{journal}{Phys. Rev. Lett.} \textbf{\bibinfo{volume}{90}},
  \bibinfo{eid}{177001}  (\bibinfo{year}{2003});
  \bibinfo{author}{\bibfnamefont{K.}~\bibnamefont{Deguchi}} et al.,
  \bibinfo{journal}{ibid. } \textbf{\bibinfo{volume}{92}},
  \bibinfo{eid}{047002}  (\bibinfo{year}{2004}).


\bibitem[{\citenamefont{Aoki et~al.}(2004)\citenamefont{Aoki, Sakakibara,
  Shishido, Settai, Onuki, Miranovi\'{c}, and Machida}}]{HAoki:2004}
\bibinfo{author}{\bibfnamefont{H.}~\bibnamefont{Aoki}} et al.,
  \bibinfo{journal}{J. Phys.: Cond. Mat.}
  \textbf{\bibinfo{volume}{16}}, \bibinfo{pages}{L13} (\bibinfo{year}{2004}).

\bibitem{FYu:1995} F.Yu et al., Phys. Rev. Lett. {\bf 74}, 5136
(1995); H.~Aubin et al., ibid. {\bf 78}, 2624 (1997).

\bibitem[{\citenamefont{Izawa et~al.}(2001{\natexlab{a}})\citenamefont{Izawa,
  Takahashi, Yamaguchi, Matsuda, Suzuki, Sasaki, Fukase, Yoshida, Settai, and
  Onuki}}]{KIzawa:other}
\bibinfo{author}{\bibfnamefont{K.}~\bibnamefont{Izawa}} et al.,
  \bibinfo{journal}{Phys. Rev. Lett.} \textbf{\bibinfo{volume}{86}},
  \bibinfo{pages}{2653 } (\bibinfo{year}{2001}{\natexlab{a}});
  \bibinfo{journal}{} \textbf{\bibinfo{volume}{88}},
  \bibinfo{pages}{027002} (\bibinfo{year}{2002}{\natexlab{a}});
  \bibinfo{journal}{} \textbf{\bibinfo{volume}{89}},
  \bibinfo{pages}{137006} (\bibinfo{year}{2002}{\natexlab{b}});
  \bibinfo{journal}{} \textbf{\bibinfo{volume}{90}},
  \bibinfo{pages}{117001} (\bibinfo{year}{2003});
  \bibinfo{author}{\bibfnamefont{T.}~\bibnamefont{Watanabe}} et al.,
  \bibinfo{journal}{Phys. Rev. B} \textbf{\bibinfo{volume}{70}},
  \bibinfo{pages}{184502} (\bibinfo{year}{2004}{\natexlab{a}})

\bibitem[{\citenamefont{Izawa et~al.}(2001{\natexlab{b}})\citenamefont{Izawa,
  Yamaguchi, Matsuda, Shishido, Settai, and Onuki}}]{KIzawa:CeCoIn5}
\bibinfo{author}{\bibfnamefont{K.}~\bibnamefont{Izawa}} et al.,
  \bibinfo{journal}{Phys. Rev. Lett.} \textbf{\bibinfo{volume}{87}},
  \bibinfo{pages}{057002} (\bibinfo{year}{2001}{\natexlab{b}}).




\bibitem[{\citenamefont{Volovik}(1993)}]{GVolovik:1993}
\bibinfo{author}{\bibfnamefont{G.~E.} \bibnamefont{Volovik}},
  \bibinfo{journal}{JETP Letters} \textbf{\bibinfo{volume}{58}},
  \bibinfo{pages}{469 } (\bibinfo{year}{1993}).

\bibitem[{\citenamefont{K{\"u}bert and Hirschfeld}(1998)}]{CKubert:1998}
\bibinfo{author}{\bibfnamefont{C.}~\bibnamefont{K{\"u}bert}} \bibnamefont{and}
  \bibinfo{author}{\bibfnamefont{P.~J.} \bibnamefont{Hirschfeld}},
  \bibinfo{journal}{Phys. Rev. Lett.} \textbf{\bibinfo{volume}{80}},
  \bibinfo{pages}{4963 } (\bibinfo{year}{1998}); I.~Vekhter and
  P.~J.~Hirschfeld, Physica C {\bf 341}-{\bf 348}, 1947 (2000);
  P.~Thalmeier and K.~Maki, Phys Rev. B {\bf 67}, 092510 (2003).

\bibitem[{\citenamefont{Vekhter and Houghton}(1999)}]{IVekhter:1999}
\bibinfo{author}{\bibfnamefont{I.}~\bibnamefont{Vekhter}} \bibnamefont{and}
  \bibinfo{author}{\bibfnamefont{A.}~\bibnamefont{Houghton}},
  \bibinfo{journal}{Phys. Rev. Lett.} \textbf{\bibinfo{volume}{83}},
  \bibinfo{pages}{4626} (\bibinfo{year}{1999}).

\bibitem[{\citenamefont{Durst et~al.}(2003)\citenamefont{Durst, Vishwanath, and
  Lee}}]{ACDurst:2003}
\bibinfo{author}{\bibfnamefont{A.~C.} \bibnamefont{Durst}},
  \bibinfo{author}{\bibfnamefont{A.}~\bibnamefont{Vishwanath}},
  \bibnamefont{and} \bibinfo{author}{\bibfnamefont{P.~A.} \bibnamefont{Lee}},
  \bibinfo{journal}{Phys. Rev. Lett.} \textbf{\bibinfo{volume}{90}},
  \bibinfo{eid}{187002}  (\bibinfo{year}{2003}).


\bibitem[{\citenamefont{Kusunose}(2004)}]{HKusunose:2004}
\bibinfo{author}{\bibfnamefont{H.}~\bibnamefont{Kusunose}},
  \bibinfo{author}{\bibfnamefont{T.}~\bibnamefont{Rice}}, \bibnamefont{and}
  \bibinfo{author}{\bibfnamefont{M.}~\bibnamefont{Sigrist}},
  \bibinfo{journal}{Phys. Rev. B} \textbf{\bibinfo{volume}{66}},
  \bibinfo{pages}{214503} (\bibinfo{year}{2002});
  \bibinfo{author}{\bibfnamefont{H.}~\bibnamefont{Kusunose}},
  \bibinfo{journal}{ibid.} \textbf{\bibinfo{volume}{70}},
  \bibinfo{eid}{054509}  (\bibinfo{year}{2004}).

\bibitem[{\citenamefont{Vorontsov and Vekhter}()}]{AVorontsov:2006}
\bibinfo{author}{\bibfnamefont{A.~B.} \bibnamefont{Vorontsov}}
  \bibnamefont{and} \bibinfo{author}{\bibfnamefont{I.}~\bibnamefont{Vekhter}},
  \bibinfo{note}{(unpublished)}.

\bibitem[{\citenamefont{Luk'yanchuk and Mineev}(1987)}]{ILukyanchuk:1987}
\bibinfo{author}{\bibfnamefont{I.~A.} \bibnamefont{Luk'yanchuk}}
  \bibnamefont{and} \bibinfo{author}{\bibfnamefont{V.~P.}
  \bibnamefont{Mineev}}, \bibinfo{journal}{Zh. Eksp. i Teor. Fiz.}
  \textbf{\bibinfo{volume}{93}}, \bibinfo{pages}{2030} (\bibinfo{year}{1987}),
  \bibinfo{note}{[Sov. Phys. JETP {\bf 66}, 1168 (1987)]}.

\bibitem[{\citenamefont{Adachi et~al.}(2005)\citenamefont{Adachi,
  Miranovi\'{c}, Ichioka, and Machida}}]{HAdachi:2005}
\bibinfo{author}{\bibfnamefont{H.}~\bibnamefont{Adachi}} et al.,
  \bibinfo{journal}{Phys. Rev. Lett.} \textbf{\bibinfo{volume}{94}},
  \bibinfo{eid}{067007} (\bibinfo{year}{2005}).

\bibitem[{\citenamefont{Eilenberger}(1968)}]{GEilenberger:1968}
\bibinfo{author}{\bibfnamefont{G.}~\bibnamefont{Eilenberger}},
  \bibinfo{journal}{Z. Phys.} \textbf{\bibinfo{volume}{214}},
  \bibinfo{pages}{195} (\bibinfo{year}{1968});
  \bibinfo{author}{\bibfnamefont{A.~I.} \bibnamefont{Larkin}} \bibnamefont{and}
  \bibinfo{author}{\bibfnamefont{Y.~N.} \bibnamefont{Ovchinnikov}},
  \bibinfo{note}{Sov. Phys.
  JETP {\bf 28}, 1200(1969)}.


\bibitem[{\citenamefont{Brandt et~al.}(1967)\citenamefont{Brandt, Pesch, and
  Tewordt}}]{BPT:1967}
\bibinfo{author}{\bibfnamefont{U.}~\bibnamefont{Brandt}},
  \bibinfo{author}{\bibfnamefont{W.}~\bibnamefont{Pesch}}, \bibnamefont{and}
  \bibinfo{author}{\bibfnamefont{L.}~\bibnamefont{Tewordt}},
  \bibinfo{journal}{Z. Phys.} \textbf{\bibinfo{volume}{201}},
  \bibinfo{pages}{209 } (\bibinfo{year}{1967});
  \bibinfo{author}{\bibfnamefont{W.}~\bibnamefont{Pesch}}, \bibinfo{journal}{Z.
  Phys. B} \textbf{\bibinfo{volume}{21}}, \bibinfo{pages}{263 }
  (\bibinfo{year}{1975}).


\bibitem[{\citenamefont{Brandt}(1976)}]{EHBrandt:1976}
\bibinfo{author}{\bibfnamefont{E.~H.} \bibnamefont{Brandt}},
  \bibinfo{journal}{J. Low Temp. Phys.} \textbf{\bibinfo{volume}{24}},
  \bibinfo{pages}{409 } (\bibinfo{year}{1976}).

\bibitem{comment1} The BPT method gives the correct
$H=0$ limit \cite{AHoughton:1998}. However, since averaging over
the intervortex distance ($\sim\Lambda$) prior to impurity
averaging is allowed only when $\Lambda/l\ll 1$, the approach does
breaks down at low $H$.

\bibitem[{\citenamefont{Won and Maki}(1996)}]{HWon:1996}
\bibinfo{author}{\bibfnamefont{H.}~\bibnamefont{Won}} \bibnamefont{and}
  \bibinfo{author}{\bibfnamefont{K.}~\bibnamefont{Maki}},
  \bibinfo{journal}{Phys. Rev. B} \textbf{\bibinfo{volume}{53}},
  \bibinfo{pages}{5927 } (\bibinfo{year}{1996}).

\bibitem[{\citenamefont{Tewordt and Fay}(2005)}]{LTewordt:2005}
\bibinfo{author}{\bibfnamefont{L.}~\bibnamefont{Tewordt}} \bibnamefont{and}
  \bibinfo{author}{\bibfnamefont{D.}~\bibnamefont{Fay}},
  \bibinfo{journal}{Phys. Rev. B} \textbf{\bibinfo{volume}{72}},
  \bibinfo{eid}{014502}  (\bibinfo{year}{2005}).

\bibitem[{\citenamefont{Houghton and Vekhter}(1998)}]{AHoughton:1998}
\bibinfo{author}{\bibfnamefont{A.}~\bibnamefont{Houghton}} \bibnamefont{and}
  \bibinfo{author}{\bibfnamefont{I.}~\bibnamefont{Vekhter}},
  \bibinfo{journal}{Phys. Rev. B} \textbf{\bibinfo{volume}{57}},
  \bibinfo{pages}{10831} (\bibinfo{year}{1998}).


\bibitem[{\citenamefont{Hirschfeld et~al.}(1988)\citenamefont{Hirschfeld,
  W{\"o}lfle, and Einzel}}]{PJHirschfeld:1988}
\bibinfo{author}{\bibfnamefont{P.~J.} \bibnamefont{Hirschfeld}},
  \bibinfo{author}{\bibfnamefont{P.}~\bibnamefont{W{\"o}lfle}},
  \bibnamefont{and} \bibinfo{author}{\bibfnamefont{D.}~\bibnamefont{Einzel}},
  \bibinfo{journal}{Phys. Rev. B} \textbf{\bibinfo{volume}{37}},
  \bibinfo{pages}{83 } (\bibinfo{year}{1988}).

\bibitem[{\citenamefont{Graf et~al.}(1996)\citenamefont{Graf, Yip, Sauls, and
  Rainer}}]{MJGraf:1996}
\bibinfo{author}{\bibfnamefont{M.~J.} \bibnamefont{Graf}} et al.,
  \bibinfo{journal}{Phys. Rev. B} \textbf{\bibinfo{volume}{53}},
  \bibinfo{pages}{15147} (\bibinfo{year}{1996}).

\bibitem[{\citenamefont{Klimesch and Pesch}(1978)}]{PKlimesch:1978}
\bibinfo{author}{\bibfnamefont{P.}~\bibnamefont{Klimesch}} \bibnamefont{and}
  \bibinfo{author}{\bibfnamefont{W.}~\bibnamefont{Pesch}}, \bibinfo{journal}{J.
  Low Temp. Phys.} \textbf{\bibinfo{volume}{32}}, \bibinfo{pages}{869 }
  (\bibinfo{year}{1978}).

\bibitem{comment2} We checked that moderate changes in the shape of the
Fermi surface and in the scattering rate lead to small
quantitative changes in the amplitude of the effects discussed
here, but do not alter any of our conclusions
\cite{AVorontsov:2006}.

\bibitem[{\citenamefont{Vekhter et~al.}(2001)\citenamefont{Vekhter, Hirschfeld,
  and Nicol}}]{IVekhter:2001}
\bibinfo{author}{\bibfnamefont{I.~}~\bibnamefont{Vekhter}},
  \bibinfo{author}{\bibfnamefont{P.~J.~}~\bibnamefont{Hirschfeld}},
  \bibnamefont{and} \bibinfo{author}{\bibfnamefont{E.~J.~}~\bibnamefont{Nicol}},
  \bibinfo{journal}{Phys. Rev. B} \textbf{\bibinfo{volume}{64}},
  \bibinfo{pages}{064513} (\bibinfo{year}{2001}).


\bibitem{PMiranovic:2003}
    Similar behavior of the residual DOS at $T=0$ was found in
\bibinfo{author}{\bibfnamefont{P.} \bibnamefont{Miranovi{\'c} et al.}},
  \bibinfo{journal}{Phys. Rev. B} \textbf{\bibinfo{volume}{68}},
  \bibinfo{pages}{052501} (\bibinfo{year}{2003}); M.~Udagawa, Y.~Yanase, and M.~Ogata,
  {\it ibid.} {\bf 70}, 184515 (2004), but not
  separated from the in-plane anisotropy of $H_{c2}$. We verified that the
  behavior of $N(0,\bm H)$ is the same for isotropic
  $H_{c2}$, and hence is intrinsic.

\bibitem{T0} The value of $T_0$ weakly depends on the shape of the Fermi
surface and the impurity scattering rate.

\bibitem[{\citenamefont{Bianchi et~al.}(2003)\citenamefont{Bianchi, Movshovich,
  Capan, Pagliuso, and Sarrao}}]{bia03}
\bibinfo{author}{\bibfnamefont{A.}~\bibnamefont{Bianchi}} et al.
  \bibinfo{journal}{Phys. Rev. Lett.} \textbf{\bibinfo{volume}{91}},
  \bibinfo{pages}{187004} (\bibinfo{year}{2003}). Here we study only the regime far from the
  FFLO state.




\bibitem{KMaki:1967}
\bibinfo{author}{\bibfnamefont{K.} \bibnamefont{Maki}},
  \bibinfo{journal}{Phys. Rev.} \textbf{\bibinfo{volume}{158}},
  \bibinfo{pages}{397} (\bibinfo{year}{1967}).


\end{thebibliography}

\end{document}